\documentclass[journal=jpccck,manuscript=article,layout=twocolumn,10pt,letterpaper]{achemso}

\usepackage[version=3]{mhchem} % Formula subscripts using \ce{}
\usepackage{mathrsfs} %\mathscr{R} and \mathscr{R}
\usepackage{amsmath,amssymb}
\usepackage{siunitx}

\author{Robert G. West}
\author{Kostas Kanellopulos}
\author{Silvan Schmid}
\email{silvan.schmid@tuwien.ac.at}
\affiliation[TU Wien]
{Institute of Sensor and Actuator Systems, TU Wien, Gusshausstrasse 27-29, 1040 Vienna, Austria.}

\title{Photothermal Microscopy \& Spectroscopy with Nanomechanical Resonators}

\abbreviations{NEMS, IR, UV}

\begin{document}

\begin{abstract}
In nanomechanical photothermal absorption spectroscopy and microscopy, the measured substance becomes a part of the detection system itself, inducing a nanomechanical resonance frequency shift upon thermal relaxation. Suspended, nanometer-thin ceramic or 2D material resonators are innately highly-sensitive thermal detectors of localized heat exchanges from substances on their surface or integrated into the resonator itself. Consequently, the combined nanoresonator-analyte system is a self-measuring spectrometer and microscope: responding to a substance’s transfer of heat over the entire spectrum for which it absorbs, according to the intensity it experiences. Limited by their own thermostatistical fluctuation phenomena, nanoresonators have demonstrated sufficient sensitivity for measuring trace analyte as well as single particles and molecules with incoherent light or focused and unfocused coherent light. They are versatile in their design, support various sampling methods and hyphenation with other spectroscopic methods, and are capable in a wide range of applications including fingerprinting, separation science, and surface sciences.
\end{abstract}

\section{Introduction}\label{introduction}
%State-of-the-art absorption spectroscopy
Absorption spectroscopy is employed in species identification and investigations into key properties, function, and behavior of molecules and nanoparticles. Traditionally, absorption spectroscopy has been performed on molecular ensembles, where degrees of freedom, species variability, such as isomerization, and interaction with the environment screen the intrinsic, response of the individual molecules. In heterogeneous samples, there can be significant variations in the size, shape, and composition of the particles, affecting their properties and, thus, their function and behavior in their environment. This motivates research in the absorption response at the single molecule and single nanoparticle level. Developments in photothermal spectroscopy have facilitated the measurement of ever-lessening analyte concentrations as well as sub-diffraction-limited localization of single nanoparticles and molecules.\cite{Adhikari2020,Adhikari2022,Shimizu2022} As a result, photothermal spectroscopy has transcended the capabilities of contemporary absorption spectroscopy techniques to provide a more detailed understanding of heterogeneous samples, with significant potential applications in a wide range of fields, including material science, catalysis, nanotechnology, and biophysics.

%State-of-the-art classical photothermal spectroscopy
Photothermal absorption spectroscopy incorporates a wide-ranging set of methods based upon the measurement of a substance's wavelength-dependent absorption by way of its thermal energy transfer to its environment upon relaxation. In general, the advantages of photothermal spectroscopy have been duly established: it is a label-free approach, unaffected by scattered light, and a direct measurement of the sample's absorption, whose cross-section scales with the first power of a particle's volume, as compared to the second order in scattering.\cite{Bialkowski2019,van2006absorption} This perspective article addresses a particular fledging, photothermal method, which relies on the thermally-induced mechanical resonance detuning in nanoresonators.

In contrast to the contemporary thermo-optical methods, which can be performed in liquid, matrix, or host cell, the nanomechanical frequency-detuning method does not require a thermo-refractive medium and must be performed in a gaseous or low pressure environments, typically of $<10^{-3}$ mbar. A vacuum lessens deleterious air damping and convective heat transfer, allowing for a sensitive detection paradigm for numerous applications, some exclusive to low pressure environments. Optical complexities, such as precise optical overlap, modulation of a heating laser are requirements often considered synonymously with photothermal spectroscopy. Thermo-optical studies regularly require precise focal depth of a probing laser relative to a heating laser as well as diffraction, screening, and filtering of light. However, with nanomechanical photothermal sensing, the "substrate" on which the sample is placed acts as the thermal detector. Hence, it does not share these requirements for its basic function, and is siutable for focused or unfocused, coherent or incoherent light, to spectroscopically analyze thin films,\cite{CasciCeccacci2019,Samaeifar2019} 2D materials,\cite{Kirchhof2022,Kirchhof2023} surface-adsorbed chemical species,\cite{biswas2014femtogram,Kurek2017,Luhmann2023} explosives,\cite{biswas2014femtogram} nanoparticle ensembles,\cite{Larsen2013,Yamada2013,Andersen2016} and individual single nanoparticles.\cite{ramos2018nanomechanical,Rangacharya2020,kanellopulos2023}. It is worth noting, that micro-fabrication methods used to make nanoresonators have also inspired suspended microchannel resonators (SMRs), which enable photothermal resonance-tuning detection in microfluidics. SMRs are capable of single cell and single protein mass detection and have duly demonstrated sensitivities in hundreds of nJ for absorption spectroscopy in the static, surface-stress tuning mode and also in frequency-tuning, dynamic mode.\cite{Pastina2020,Miriyala2016} However, the primary thrust of this perspective is to introduce a wider community to the accomplishments and direction of research in nanomechanical resonators on the path of ever-increasing sensitivity toward single-molecule absorption spectromicroscopy.

\begin{figure}
\centering
\includegraphics[width=0.475\textwidth]{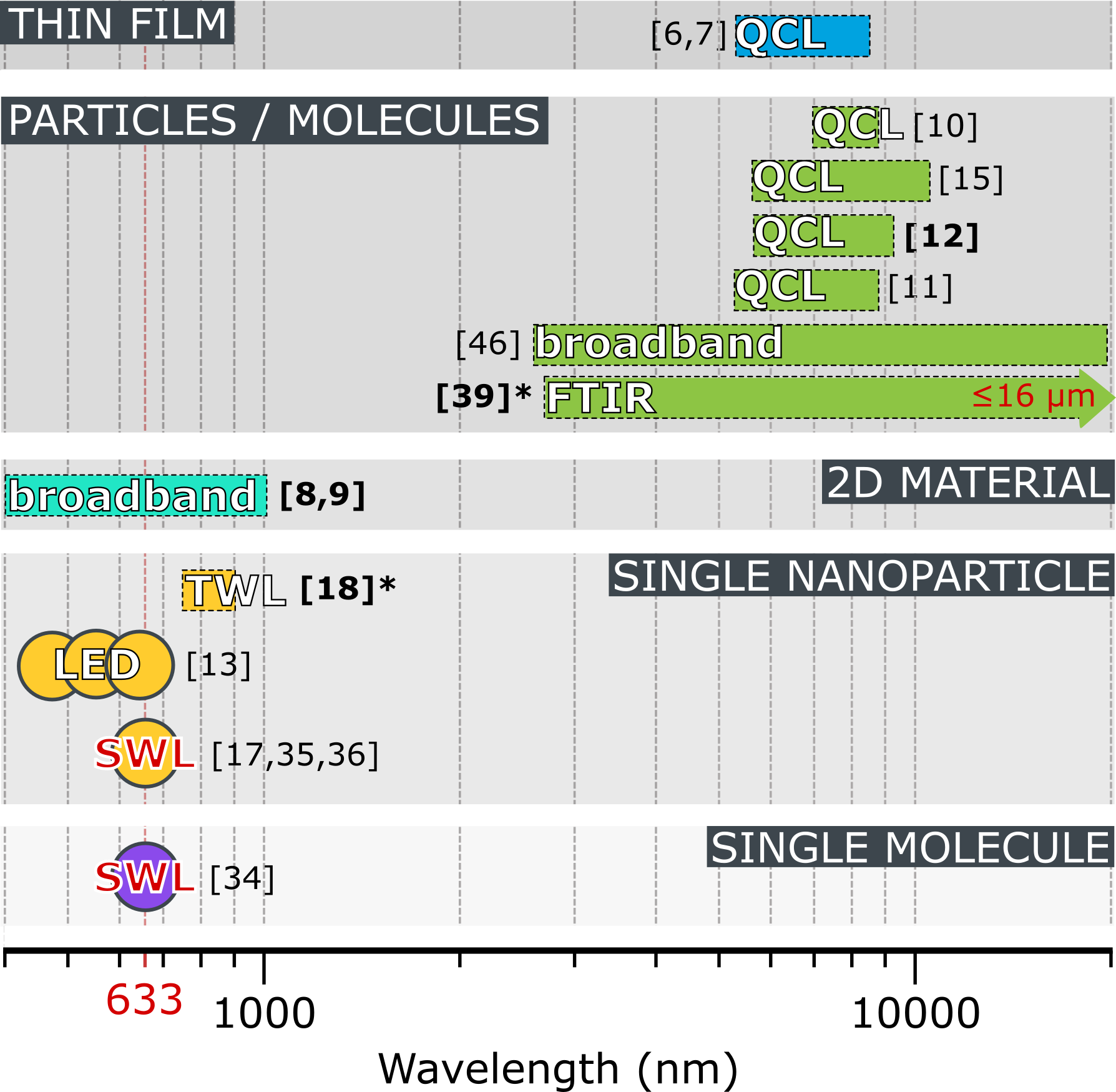}
\caption{Nanomechanical resonators are a platform for spectroscopy. The collection of studies so far demonstrate their capability for measuring a broad range of substances down to single particles and molecules by photothermal spectroscopy. The bold references in brackets represent studies in the last two years; those marked with an asterisk have not yet been published. The only limitation of the spectral bandwidth is the absorption spectrum of the measured substance itself. This fact is evidenced by the numerous methods of illumination: a single-wavelength laser (SWL), light-emitting diode (LED), tunable wavelength laser (TWL), quantum cascade lasers (QCL), broadband light (such as from a lamp) transmitted through a monachromator, or FTIR.}
\label{Fig:StudyComparison}
\end{figure}

%from deflection to resonant photothermal sensing
Though explorations into NEMS have led down many avenues of cutting-edge research, only tens of applications showcase their potential for photothermal spectroscopy (see Figure \ref{Fig:StudyComparison}), but greater consideration is being given to this research as of late. Nonetheless, the concept of spectroscopy by changes in stress in a resonant structure are not new, having been introduced in 1969 for the design of nanomechanical thermal infrared detectors.\cite{cary1969} The concept of photothermal spectroscopy with microstructures began with bimaterial cantilevers, whose \qty{100}{\pico\watt} thermal sensitivity, dependent upon quasi-static thermal stress-induced bending, already showed great promise in the early 1990s.\cite{Barnes1994,Barnes1994a} %One vein of research has focused on improvements in this static mode concept.\cite{Yun2011,Kim2013,Bagheri2014,FaheemKhan2014,Kim2019} 
Due to the relatively long thermal time constants, these quasi-static deflection measurements have to be performed at low frequencies of a few tens to hundreds of Hz, a frequency range that typically is strongly affected by 1/f noise.\cite{varesi1997photothermal} Conversely, nanomechanical resonators, such as strings, drumheads, or trampolines, are highly sensitive to temperature variations, as seen in the shift of their resonant frequency. The higher operation frequencies in the kHz to MHz regime is one of the big advantages of photothermal sensing with nanomechanical resonators.

\begin{figure*}
\centering
\includegraphics[width=0.8\textwidth]{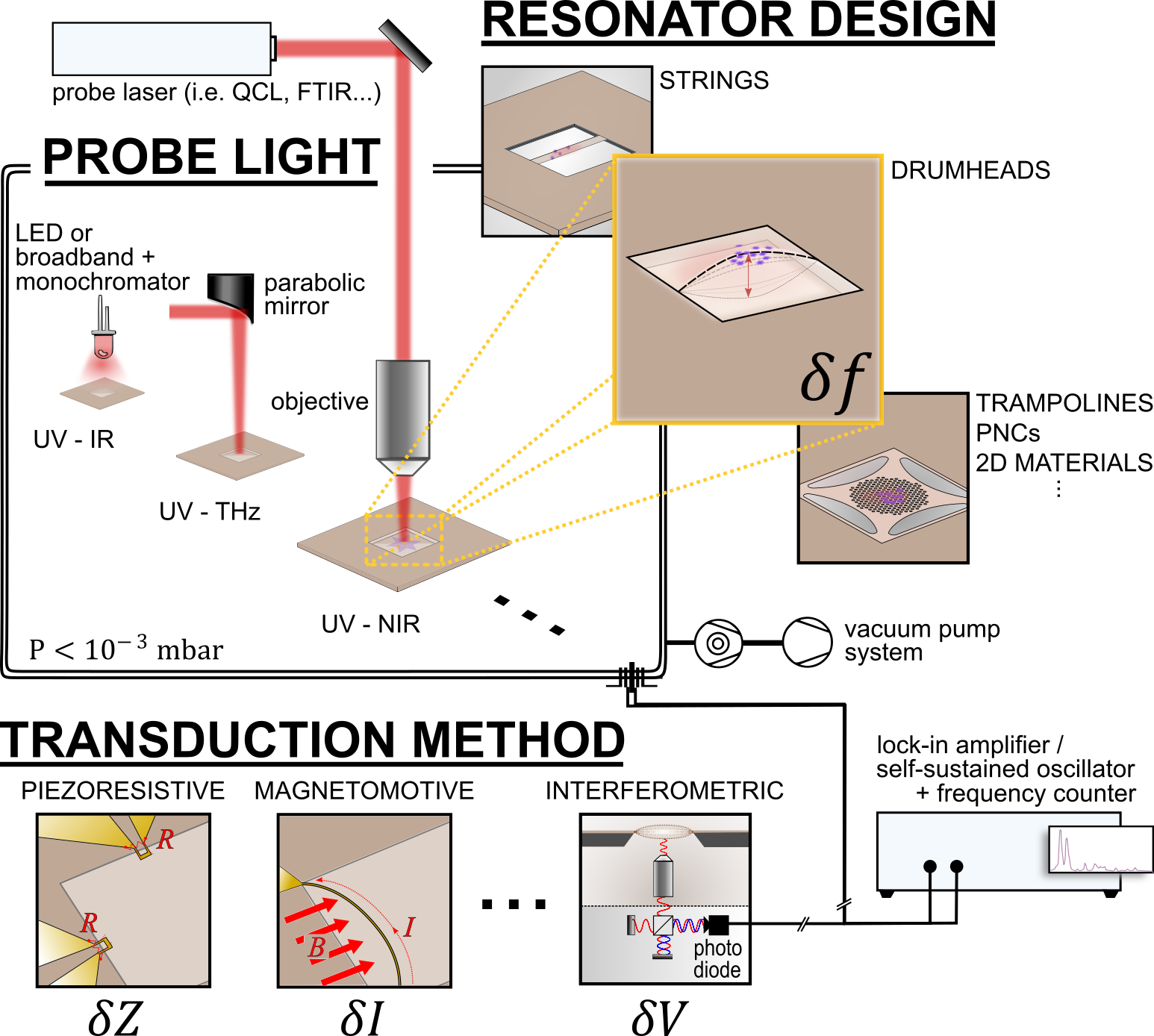}
\caption{Nanomechanical resonators are highly versatile and adaptable in design and function, to perform highly-sensitive broadband absorption spectroscopy for numerous applications in low-pressure environments. They also avail themselves to various means of transduction and frequency-tracking schemes.}
\label{Fig:PhotothermScheme}
\end{figure*}

%Explanation of the underlying mechanics
The fundamental principle, which allows nanomechanical systems to be highly sensitive spectrometers is straightforward: light, which is absorbed by the sample, anywhere from UV to THz, is dissipated as heat into the thin resonator upon thermal relaxation. This induces a frequency shift, measured optically by interferometry or electrically, by a variety of methods (see Figure \ref{Fig:PhotothermScheme}). Resonating, nanometer-thick microstructures, such as strings, drumheads, or trampolines, are exceptionally sensitive to local changes in temperature, inducing a change in tensile stress and ultimately limited by fundamental, statistical thermal fluctuations including that described by the fluctuation dissipation theorem.\cite{vig1999noise,Cleland2002} Though the weight and interplay of these manifestations in the frequency noise is still an ongoing investigation;\cite{Zhang2023demonstration,Sadeghi2020a,piller2020thermal,zhang2020radiative} the extent of heat dissipation and thermal interfaces in these thin, suspended structures are clearly definable, as in the case of a point-source heating the resonator, which will be introduced in the discussion of Fundamental Principles \& Limitations\ref{sec:FPL}. To this end, the ultimate goal is to operate nanomechanical photothermal sensors at the ultimate thermal noise limit to create, on the one hand, broadband detectors for electromagnetic radiation \cite{Blaikie2019,piller2022thermal,Zhang2013,qian2019high}, and on the other hand, single-molecule absorption spectromicroscopes.

\begin{figure*}
\centering
\includegraphics[width=\textwidth]{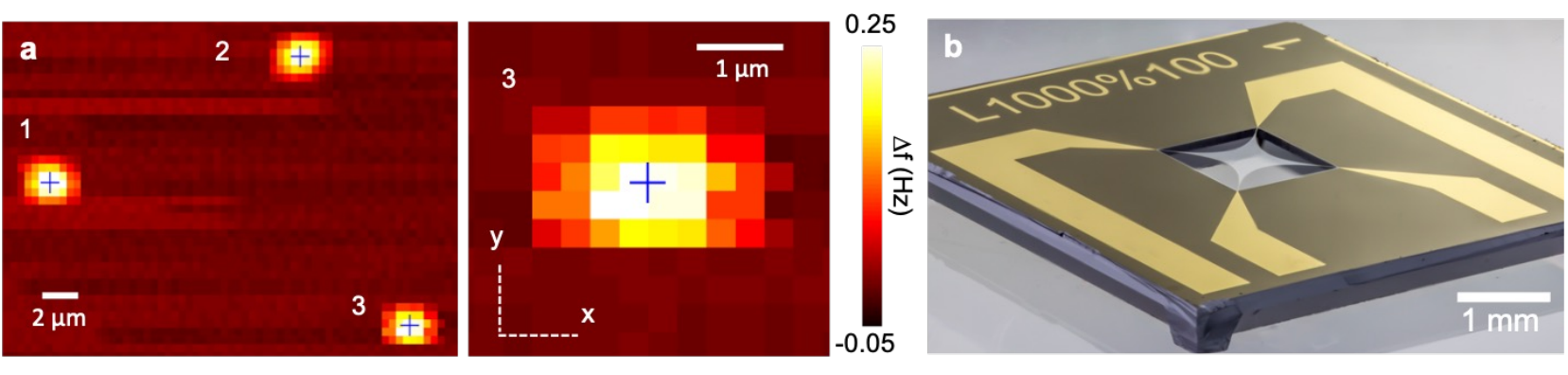}
\caption{Nanomechanical resonators are a platform for microscopy. (a) Drumhead nanoresonator frequency detuning due to photothermal heating of three individual molecules (Atto633, a fluorescent dye) from a raster scan of the excitation laser at a wavelength of 633 nm with a signal-to-noise ratio of $\sim$70 (Reprinted from ref. \citenum{Chien2018}). (b) Photograph of a trampoline-shaped nanomechanical SiN resonator similar to the one used for localizing 50 nm-long gold nanorods and probing their polarization dependence as seen in Fig. 4a (ref. \citenum{Chien2020}).}
\label{Fig:Chien18_atto633scan}
\end{figure*}

\begin{figure*}
\centering
\includegraphics[width=1\textwidth]{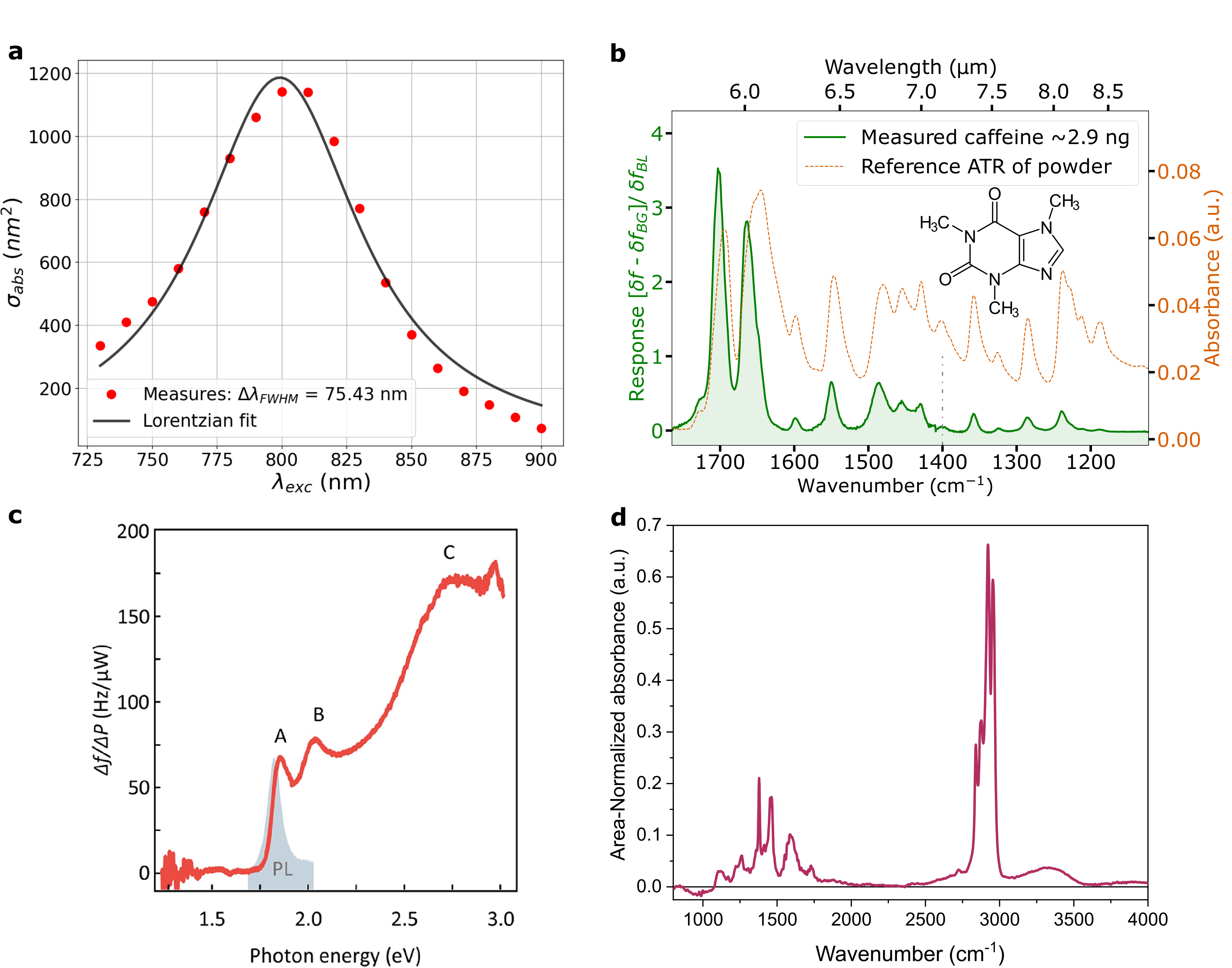}
\caption{Nanomechanical photothermal absorption spectra from the most recent works demonstrating the wavelength range and variability of probing source. (a) Near-IR plasmon absorption spectrum of a 50 nm-long silica-coated Au nanorod, using a tunable-wavelength laser (TWL), associated with measurements in ref. \citenum{kanellopulos2023} (unpublished). (b) Comparison of IR spectra of aerosol impacted caffeine by nanomechanical photothermal spectroscopy with that of the FTIR-ATR spectrum; the probe is a quantum cascade laser (ref. \citenum{Luhmann2023}). (c) Raw frequency response of a MoS2-SiN-hybrid resonator device as a function of photon energy using a TWL (ref. \citenum{Kirchhof2022}). (d) Initial findings of an estimated 124 ng of polypropylene nanoparticles absorption using a SiN trampoline resonator using the principle of FTIR performed by \citenum{ILLabs} (unpublished). Data in (a) and (d) are not intended for publication elsewhere.}
\label{Fig:MultipleSpectra}
\end{figure*}

% Photothermal microscopy and localization
Focused light enables photothermal microscopy in addition to spectroscopy, localizing particles far below the diffraction limit and allowing for their characterization. Individual, fluorescing Atto633 molecules have been localized on SiN drumhead resonators by Chien, et. al (see Figure \ref{Fig:Chien18_atto633scan}a).\cite{Chien2018} With SiN trampolines (Figure \ref{Fig:Chien18_atto633scan}b), the same authors spin-coated 200 nm diameter gold nanoparticles and subsequently positioned them with the aid of an atomic force microscopy (AFM) tip.\cite{Chien2020} Then, they were localized photothermally with a 633 nm laser with 3 $\AA$ resolution. Likewise, trampoline resonators verified the polarization dependence of the photothermal response of single $\sim$50 nm-long gold nanorods and revealed their specific absorption spectra (see Figure \ref{Fig:MultipleSpectra}a).\cite{Chien2020,kanellopulos2023} Nano-trampolines also enabled analysis of carbon content of direct-write plasmonic Au nanostructures and verification of enhancement localized surface plasmon fields between two direct-write bowtie structures.\cite{Chien2021} Metal nanoparticles with various other shapes were probed on SiN strings.\cite{Schmid2014a,Rangacharya2020} The nanoresonators' response to thermal transfer from the plasmonically heated particles allowed the authors to obtain precise values of their absorption cross-sections. %localizing below the diffraction limit is different than resolving below the diffraction limit

% Photothermal IR spectroscopy
Beyond visible wavelengths, ventures into the infrared (IR) spectrosocpy began with absorbed trace concentrations of explosive mixtures, identified by their spectral fingerprints in the lower IR region. In a pilot study with less than 100 pg of explosives, absorbed light from a quantum cascade laser (QCL) induced nanomechanical bending of cantilevers upon absorption,\cite{Kim2013} inspiring further, resonant nanostring detection of 42 fg, or 190 attomoles of RDX.\cite{biswas2014femtogram} Infrared spectroscopy of nanomechanical resonators using dispersed IR light through a monochromator made its debut in a study by Yamada et. al. on SiN strings.\cite{Yamada2013} The string resonators were capable of detecting their own absorption spectrum and that of aerosol sampled polyvinylpyrrolidone (PVP). The sensitivity of the resonators was even sufficient to measure the IR spectra of organic compounds coated on $TiO_2$ nanoparticles. The estimated limit of detection was 44 fg for obtaining the IR fingerprint spectra of nanoengineered materials. Soon after,  Other studies with SiN strings affirmed their sensitivity, achieving similar signal strengths and resolutions equal to that of ATR-FTIR with far less analyte.\cite{Andersen2016} Their versatility in application was demonstrated with spectral characterization of Tadalafil drug and PVP thin films with a quantum cascade laser (QCL).\cite{CasciCeccacci2019,Samaeifar2019} The signal-to-noise ratio (SNR) for 200 nm polymer films was almost six times higher than with FTIR-ATR, and 20 nm-thick layers with a SNR of 307.

% desorption
This handful of applications indisputably demonstrates nanoresonators' capability as direct, \textit{in situ} spectrometers, whose wavelength range is only constrained by the analyte absorption spectrum. Nanomechanical resonators have shown capability for use in tandem with various optical methods, most recently including Fourier Transform Infrared Spectrometry.\cite{ILLabs} In the same vein, nanomechanical photothermal spectroscopy can be used to monitor chemiphysical processes occurring on the resonator surface or the physical behavior of the resonator itself. Some examples of such orthogonal processes are desorption and phase transitions.\cite{Shakeel2018,CasciCeccacci2019,Karl2018} Such spectral "eavesdropping" during thermal desorption was demonstrated in 2014, as the absorption spectra of a mere 2 fg (that is 190 attomoles) of RDX condensate was measured on silicon nitride nanostrings.\cite{biswas2014femtogram} A recent study by Luhmann et al. built upon this concept with thermal desorption of various mass loads of aerosol-impacted caffeine and theobromine, hyphenating NEMS-based thermal desorption with IR spectroscopy (NEMS-IR-TD).\cite{Luhmann2023} In harmony with mass sensing capability of the nanoresonator, the authors obtained characteristic thermal programmed desorption (TPD) dynamical traces of the analyte and condensates in addition to their spectra. The spectrum, as compared to FTIR-ATR spectra of caffeine powder is shown in Figure \ref{Fig:MultipleSpectra}b. With the aid of a Peltier element for temperature control, an analysis of time-dependent spectra during isothermal desorption allowed for separation of the spectra of species with differing desorption energies by time-resolved spectral analysis. This highlights the method's potential use in separation science. Along these lines, a growing body of literature is supporting the idea that the current utility of nanomechanical photothermal spectroscopy extends to studies in material physics and potentially to surface physics.

% Kirchhof's papers
Two recent, significant contributions by Kirchhof et al. bear archetypical significance toward this point. In their first study,\cite{Kirchhof2022} they obtain the \textit{in situ} absorption spectra of 2D materials, from 400 $nm$ to 1 $\mu m$ (see Figure \ref{Fig:MultipleSpectra}c), integrated into the resonator itself. These structures represent extended single molecules, or $\sim7.2\times10^8$ atoms, for which a measurement of the simultaneously reflected light is then corroborated to obtain the most precise, and conceivably accurate, determination of a 2D material's dielectric function to date. This demonstrates nanomechanical photothermal spectroscopy at the fundamental limit for 2D materials, as the substance being measured is entirely the photodetector itself. Their second study,\cite{Kirchhof2023} demonstrates spectral measurements along with a noise-equivalent power (NEP) of 890 $fW/\sqrt{Hz}$ of silicon nitride drumhead resonators, on which the 2D materials were integrated, at room temperature. This sensitivity is on the same order of the commercially-available silicon avalanche photodiodes at maximum sensitivity but over a much broader wavelength range. The hybrid 2D-material-resonator systems readily yield the excitonic transition of $WS_2$, plasmonic modes and intraband transitions of a plasmonic supercrystal, and dipole-dipole excited state transitions in $CrPS_4$. Spectra of the latter material is said to have been attained with a mere 1 $\mu{W}$ incident radiation, and the SiN drumhead suffers not even a two-fold reduction in responsivity at intermediate temperatures down to 4 $K$. %(yes, same order: 240 fW/sqrtHz, Thorlabs)

In a broader aspect, the faculty of nanomechanical photothermal sensing is also merited by its versatility of implementation in diverse applications, which can be performed in conjunction with its photothermal capabilities. This includes its full compatibility with the Fourier transform infrared absorption spectroscopy (FTIR) concept, pioneered by Invisible-Light Labs GmbH.\cite{ILLabs} Figure \ref{Fig:MultipleSpectra}d shows one such preliminary result of 124 ng of aerosol-impacted polypropylene nanoparticle dispersion, with $\sim$50 nm diameter (Lab261 PP50), measured with NEMS photothermal spectroscopy as compared to a standard ATR-FTIR spectrum.

\section{Sample Preparation on Nanomechanical Resonators}
\begin{figure*}[h]
\centering
\includegraphics[width=0.7\textwidth]{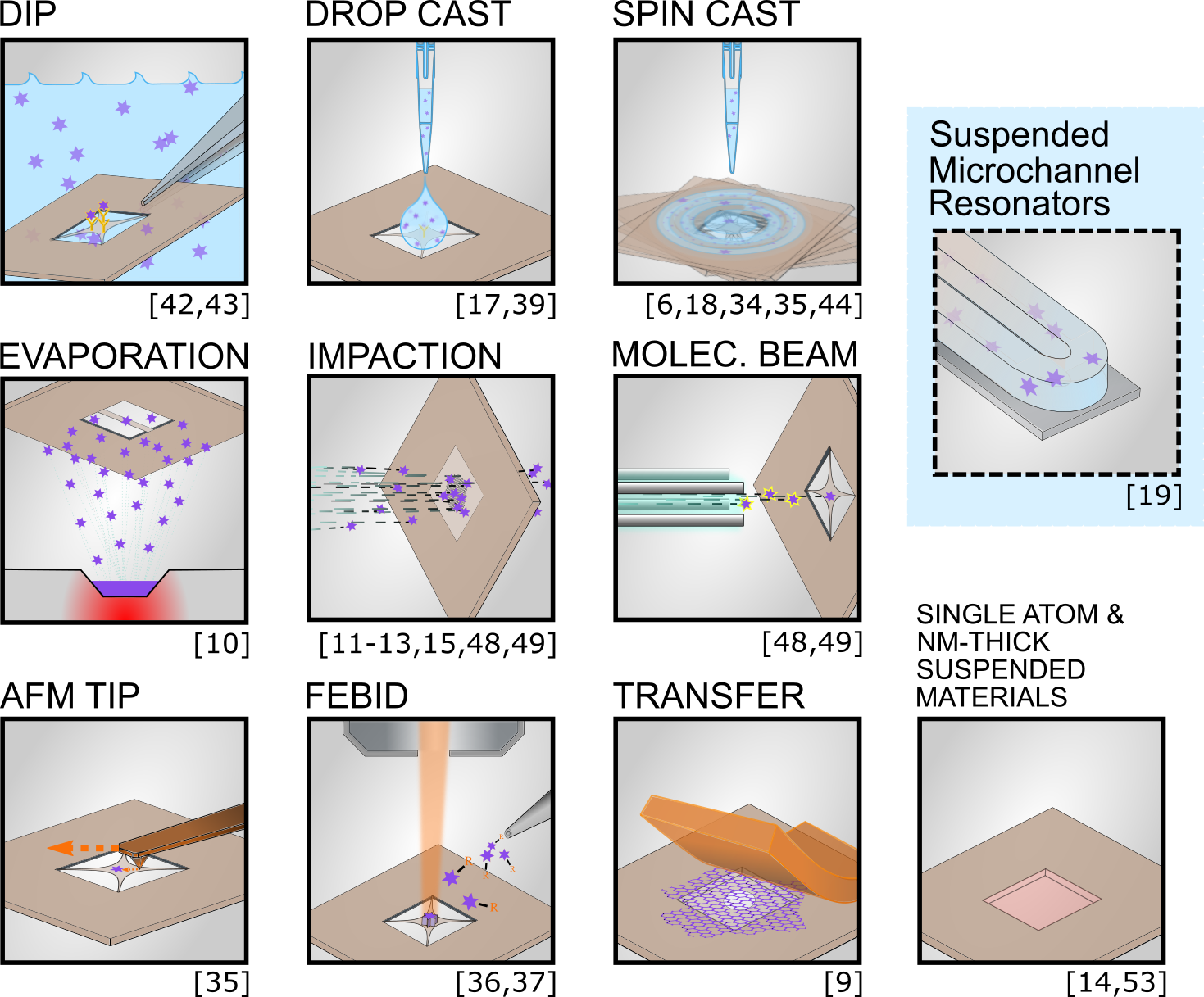}
\caption{Nanomechanical resonators are a platform for direct sampling in several mass detection, spectroscopic, and localization studies. Suspended michrochannel resonators (SMRs) rely upon the same state-of-the-art fabrication technologies as nanoresonators. Though it is considered a microresonator due to its size, it is currently the only means of directly measuring analyte in solution by the same principle as nanoresonators.}
\label{Fig:Sampling}
\end{figure*}

Nanometer-thick suspended structures facilitate a variety of sampling methods already employed in chemistry, life sciences, and surface science and materials (see Figure \ref{Fig:Sampling}). Though many of these applications do not involve spectroscopy, the resonator takes on the role of a substrate, which can subsequently be used for detection. Fabricated in their diced wafer section, or chip, the resonators can be functionalized, passivated,\cite{Luhmann2023} or left bare then dipped into a liquid suspension in the same way that microstructure detectors have been.\cite{Bagheri2014,Butt2001} Analyte can also be drop-casted or spin-coated onto the resonator-chip.\cite{Chien2018,Chien2020,Chien2020a,Rangacharya2020,CasciCeccacci2019,kanellopulos2023} Single nanoparticles, molecules, and thin films have been detected following these basic sampling procedures, and spin-coated nanoparticles can be subsequently positioned with the aid of the cantilever tip of an atomic force microscope (AFM).\cite{Chien2020} Aerosol impaction is appropriate for airborne analyte, nebulized solutions, or suspensions.\cite{Schmid2013a,Luhmann2023,Kurek2017,Yamada2013a,Andersen2016,Luhmann2023b} In vacuum, suspended nanostructures are suitable for adsorption of evaporated analyte or that produced by electrospray ionization and subsequently directed or focused into a molecular beam as in mass spectrometry.\cite{biswas2014femtogram,Naik2009,Roukes2019} For studies of the physical properties of materials, transfer by stamping or even focused electron-beam induced deposition can be considered a means of sampling.\cite{Schmid2014a,Chien2021,Kirchhof2022,Kirchhof2023}

Nanomechanical photothermal spectroscopy can, therefore, be applied to a vast range of substance distributions, from thin films to single molecules. The majority of these cases allow for pressures just below $10^{-3}$ mbar, which is achievable within minutes using a roughing pump. Though the analyte in these studies are in solid phase upon detection, an exception is found in suspended microchannel resonators (SMRs), microresonators with nanometer-thick walled channels hold pico- to femtoliters of solution (highlighted in Figure \ref{Fig:Sampling}). In this case, photothermal absorption spectroscopy in the dynamic mode, or resonant mode, follows the same principles of frequency tuning. Although the exchange of heat occurs indirectly through the solvent, this method is capable of single-particle, virus, or cell detection.\cite{Pastina2020}
%\cite{Burg2003,BryanFaheemKhan2014,Bryan2014,Miriyala2016,alodhayb2019nanomechanical}

\section{Fundamental Principles \& Limitations}
\label{sec:FPL}
When used as photothermal detectors, nanomechanical resonators are unique in that the analyte becomes a part of the detector system, coupled through thermal energy transfer. The power responsivity of the detector/substance system, having wavelength-dependent absorptance $\alpha_{abs}(\lambda)$ relates the relative, or fractional, frequency shift $y=(f_{final}-f_{initial})/f_{initial}$ to the incident power $P_0$ according to 
\begin{equation}
    \mathscr{R}_P=\frac{1}{\alpha_{abs}(\lambda)}\frac{\partial{y}}{\partial{P_0}}\,.
\end{equation}
The caveat is that, in all resonators, higher responsivity also increases sensitivity to thermal noise. The tens of studies mentioned in this perspective do not all represent geometries that have been fully optimized for the reduction of this noise, as often, their applications require compensations in the resonator design for reasons such as improving aerosol sampling efficiency. Nonetheless, a SNR of just above 400 was achievable for the strongest spectral signatures of $\sim$120 pg of indomethacin by such structures,\cite{Kurek2017} Similarly, a SNR of $\sim$337 was determined for a passivation layer, Trimethylchlorsilane, with a surface density of 0.3 fg/$\mu$m$^2$.\cite{Luhmann2023}

Regardless of these impressive signal-to-noise ratios, nanoresonators' full potential as spectrometers has not been fully explored. Though the description of noise processes in mechanical resonators and circuits are well-established\cite{vig1999noise} and adapted to nanoresonators for guiding their optimization,\cite{Cleland2002} the interplay of these processes and their dependence on the geometry of the structure is a developing discourse.\cite{piller2020thermal,Snell2022,Zhang2023} Nonetheless, an advantage of suspended nanoresonators is their endless variety of forms: strings, drumheads, trampolines, and more intricate structures, including physics-driven or topologically optimized structures such as phononic crystals and spiderwebs optimized by machine learning.\cite{Sadeghi2020,Kirchhof2020,Shin2022} Likewise, variations in fabrication, such as the reduction in stress by oxygen plasma tuning,\cite{Luhmann2017a} and in material, such as graphene,\cite{Steeneken2021} yield resonators more responsive to thermal exchanges. Other considerations include the resonator's physical interaction with the incoming light; where, for example, absorption by the resonator itself can be reduced compared to the analyte by reducing and adjusting its thickness, allowing for constructive transmission for a preferred band of wavelengths.\cite{King2012} For methods hyphenated with nanomechanical photothermal spectroscopy, ultimate sensitivity is certainly not always necessary, but applications such as single-molecule spectromicroscopy call for optimization to reduce the various forms of noise.

\subsection{Advantage of Photothermal Absorption Measurements}
\begin{figure}
\centering
\includegraphics[width=0.46\textwidth]{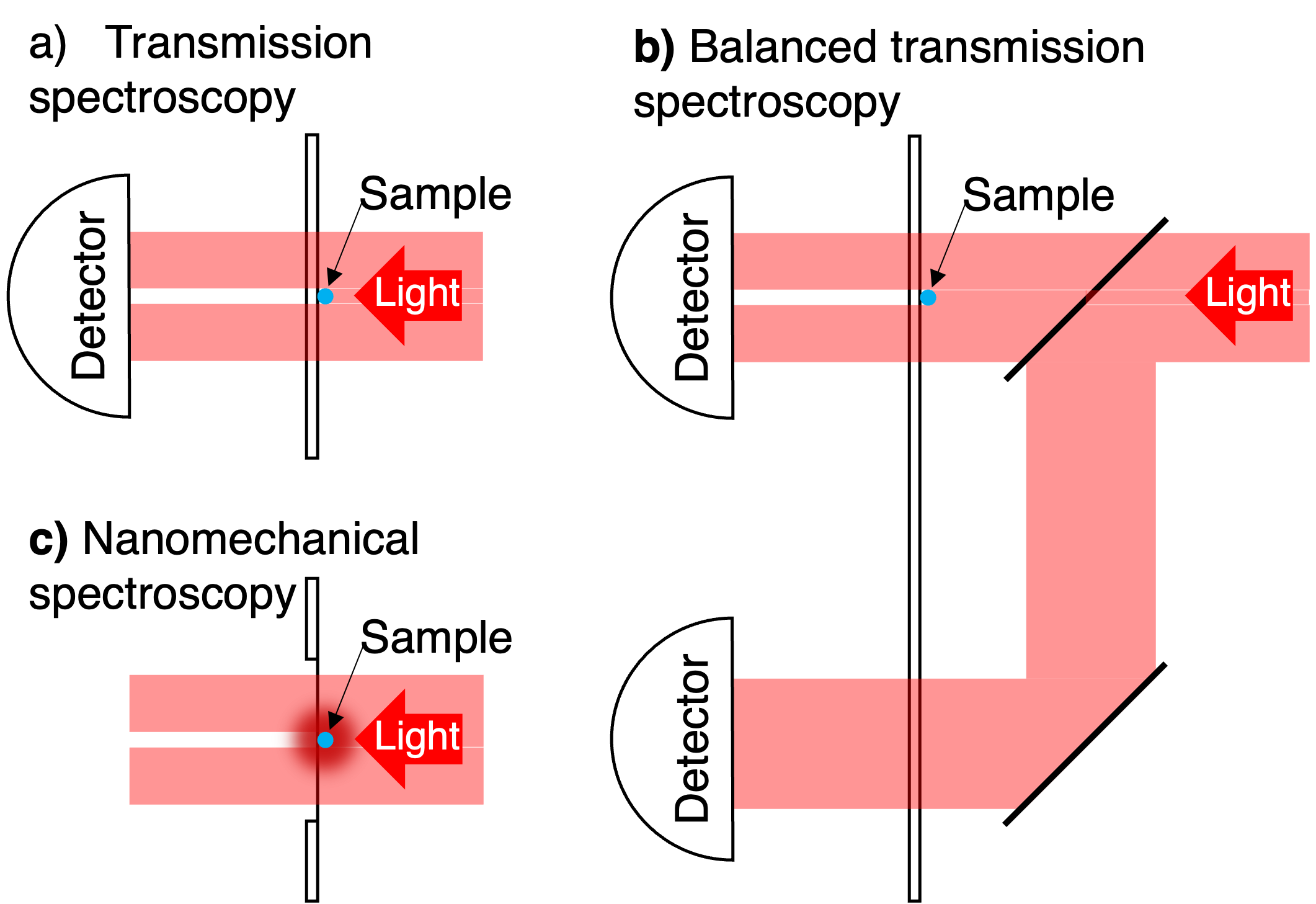}
\caption{Various means of spectroscopic analysis of a sample on a finite substrate, using a) single-ended transmission, b) balanced transmission, or c) a nanomechanical photothermal sensing scheme.}
\label{fig:T-BT-NP}
\end{figure}

%The title "advantage of absorption measurements" is not specific enough, because both transmission schemes are measurements of the absorption of the sample.
Despite the complexities, which thermal noise imparts to these highly sensitive detectors, one can appreciate photothermal spectroscopy and its superior detection strategy for analyte on, or part of, the nanomechanical resonator compared to even balanced transmission spectroscopy (see Figure \ref{fig:T-BT-NP}). Assuming a non-scattering, non-luminescent sample, the power of indicent light ($P_i$) the sample receives is equivalent to the sum of the transmitted ($P_i$), absorbed ($P_a$), and reflected ($P_r$) powers:
\begin{equation}\label{eq:powersum}
 P_i=P_t+P_a+P_r\,.
\end{equation}
This distribution of the initial power can be described relatively, by dividing equation (\ref{eq:powersum}) by the incident power. That is
\begin{equation}
 1=T+\alpha+\rho\,,
\end{equation}
where transmittance ($T=P_t/P_i$), absorptance ($\alpha=P_a/P_i$), and reflectance ($\rho=P_r/P_i$). It is important to note that absorptance should not be confused with absorbance, which is defined as the logarithm of $T$.
Typically, when analyzing small samples, the amount of absorbed light is significantly smaller than the transmitted light $T\gg\alpha$.

In transmission spectroscopy, as depicted in Figure \ref{fig:T-BT-NP}a, the sample is irradiated with probing light, and the sample's wavelength-dependent absorption is inferred from either the transmitted or the reflected light. A schematic of a transmission-based measurement setup is depicted in Figure \ref{fig:T-BT-NP}a. In this scenario, reflectance is typically much smaller than transmittance $T\gg\rho$, independent of the sensitivity of the detector that is used to measure the transmitted IR light. Assuming the probe light source has a relative intensity noise $S_I(\omega)$ with units [Hz$^{-1}$] and that all powers are measured within the same system bandwidth. Then the signal-to-noise ratio with units [Hz$^{1/2}$] for the single-ended detection scheme (Figure~\ref{fig:T-BT-NP}a) is given by 
\begin{equation}\label{eq:SNR-T}
\text{SNR}=\frac{P_a}{T\ P_i \sqrt{S_I(\omega)}}
\approx\frac{\alpha}{\sqrt{S_I(\omega)}}.
\end{equation}
For a typically small absorptance $\alpha$, this inevitably results in a small SNR. The situation gets even worse for light sources with large relative intensity noise as with IR QCLs, for example, which can be suppressed with a balanced detection scheme.\cite{akhgar2020next}

Balanced detection is a solution to eliminate correlated (non-quantum) intensity noise of the probing light source: the probing IR light is split into a probing and an identical reference beam where the light is not interacting with the sample itself. While balanced transmission measurements are technically more complex than simple transmission measurements, the resulting SNR is strongly enhanced, allowing even for the detection of single-molecule absorption. \cite{celebrano2011single} Ultimately, the SNR of a balanced detection scheme is vulnerable to quantum noise, such as shot noise.\cite{kukura2010single} A schematic of a balanced transmission scheme (Figure~\ref{fig:T-BT-NP}b) is shown in Fig.~\ref{fig:T-BT-NP}b. Subtracting the signal of the probing beam from the reference signal removes the signal due to the power fluctuations
\begin{equation}\label{eq:SNR_BT}
\begin{split}
 \text{SNR}&=\frac{P_a}{[(1-\rho)-T]\ P_i \sqrt{S_I(\omega)}}\\
 &\approx\frac{1}{\sqrt{S_I(\omega)}}.
 \end{split}
\end{equation}
This equation shows, that the SNR improves compared to (\ref{eq:SNR-T}) and it scales directly with the inverse of the relative intensity noise of the light source.

In the case of a photothermal measurement, as depicted in Fig.~\ref{fig:T-BT-NP}c, the SNR is given by
\begin{equation}\label{eq:SNR-a}
\text{SNR}=\frac{P_a}{\alpha\ P_i \sqrt{S_I(\omega)}} =\frac{1}{\sqrt{S_I(\omega)}}.   
\end{equation}
This equation shows that a photothermal measurements offer the same SNR as a balanced detection scheme (\ref{eq:SNR_BT}). In nanomechanical photothermal sensing, the sensitivity is solely dictated by the relative intensity noise of the light source, which can be due to thermal, electronic, or ultimately shot noise. Due to the typically long thermal time constant of nanomechanical resonators in the ms range, the low-frequency relative intensity noise of the light source is relevant and should be as low as possible.  

On the one hand, to fully benefit from the improved SNR (\ref{eq:SNR-a}), the  nanomechanical resonator needs to be sensitive enough to resolve the intensity fluctuations of the light source $\text{NEP} \le P_i \sqrt{S_I(\omega)}$,
where NEP is the noise equivalent power with units of [W/Hz$^{1/2}$].  On the other hand, if the sensitivity is too low to resolve the intensity fluctuations $\text{NEP} > P_i \sqrt{S_I(\omega)}$, the signal-to-noise-ratio is given by
\begin{equation}
\text{SNR} =\frac{\alpha P_i}{\text{NEP}}.
\end{equation}
The NEP of nanomechanical resonators is discussed in the next section.

\subsection{Noise-Equivalent Power}
While photon detectors can reach single photon sensitivity, the sensitivity of thermal detectors, such as bolometers or pyroelectric detectors, is typically limited by electronic noise, including Johnson noise.\cite{Rogalski2019a,lasercomponents_noise} However, mechanical thermal detectors operate differently as they do not rely on electric detection principles to detect irradiated power. Consequently, nanomechanical detectors are not limited by electronic noise, a unique feature enabling enhanced sensitivity.\cite{cary1969} %is this the correct refernce?!!

The signal from the resonator is the relative frequency shift in time, $y(t)$, and the resonator detector's noise-equivalent power is specified by a white noise spectral density ($S_y(\omega)$), and its optothermal power responsivity ($\mathscr{R}_P(\omega)$) to that noise:
\begin{equation}\label{eq:NEP}
    \text{NEP}= \frac{\sqrt{S_y(\omega)}}{\mathscr{R}_P(\omega)}\,.
\end{equation}
NEP has units of [W/Hz$^{1/2}$], since $S_y(\omega)$ is in units of [Hz$^{-1}$] and $\mathscr{R}_P$ with units [W$^{-1}$].\cite{Schmid2023}

The power responsivity, $\mathscr{R}_P(\omega)$, is an intrinsic property of a resonator, describing its frequency response to absorbed power according to
\begin{equation}\label{eq:R_P}
    \mathscr{R}_P(\omega) = \frac{\mathscr{R}_T}{G}\frac{1}{\sqrt{1+\omega^2\tau_{th}^2}}\,,
\end{equation}
where $\mathscr{R}_T$ is the fractional frequency response to a change in temperature, and $G$ is the thermal conductance. The power responsivity has a low-pass behavior, dropping for power fluctuations with frequencies faster than the resonator's thermal time constant $\tau_{th} =C_{th}/G$,
given heat capacity, $C_{th}$.

According to the definition of the NEP in equation (\ref{eq:NEP}), the sensitivity of a system can be improved by minimizing the fractional frequency noise and maximizing the responsivity. To understand this better, we look to the power responsivity (\ref{eq:R_P}), which is directly proportional to the responsivity to temperature changes $\mathscr{R}_T$ in the resonator. In general, this parameter is influenced by two resonator material parameters: the temperature-induced softening, which affects the Young's modulus, and thermal expansion, affecting the tensile stress.\cite{Schmid2023}

Specifically, in stress-free structures such as beams, plates, and cantilevers, the dominant photothermal effect is the temperature-induced softening. On the other hand, in stressed structures such as strings and drumheads, thermal expansion leading to stress reduction plays a more significant role. For a comparison of these two effects, consider a slender beam and a string, whose temperature responsivity for an even increase in temperature can then be described by $\mathscr{R}_T \approx \alpha_E/2$ and $\mathscr{R}_T \approx \alpha_{th} E/(2 \sigma)$,\cite{Schmid2023} respectively. Here, $E$ is the Youngs' modulus, $\sigma$ is the tensile stress, $\alpha_E$ is the material's temperature coefficient of Young's modulus, and $\alpha_{th}$ is the thermal expansion coefficient. Typically, the temperature and thermal expansion coefficients are of the same order of magnitude. The temperature responsivity of strings is enhanced by a factor of $E/\sigma$, which is generally much larger than unity for most materials.  Therefore, the responsivity of strings and drumheads, which can be further optimized by stress-tuning,\cite{Chien2018} is significantly larger than that of beams and plates. This enhanced responsivity makes strings and drumheads more suitable for photothermal sensing applications.

In addition to maximizing temperature responsivity, another factor that plays a role in optimizing the responsivity (\ref{eq:R_P}) is the minimization of the thermal conductance $G = G_{cond} + G_{rad}$, where the vacuum environment eliminates contributions due to convection. Here, $G_{cond}$ and $G_{rad}$ are the conductances due to conductive and radiative heat transfer, respectively. Strings and drumheads are naturally suited for achieving a high thermal isolation (small $G_{cond}$) due to their large aspect ratios (lateral size, or length to thickness). In fact, it has been demonstrated that larger silicon nitride drumheads reach a regime where radiative heat transfer dominates over conductive heat transfer ($G_{rad} > G_{cond}$).\cite{piller2020thermal,zhang2020radiative} 

According to Stefan-Boltzmann's law for small temperature variations, the radiative heat conductance is given by $G_{rad}\approx 4 A_s \varepsilon \sigma_B T^3$,\cite{vig1999noise} with the surface area $A_s$, the Stefan-Boltzmann constant $\sigma_B$, temperature $T$, and the emissivity $\varepsilon$. As a fundamental physical process, which cannot be completely eliminated, even in materials with very low thermal conductivity, radiative heat transfer establishes the lower limit of the overall achievable heat conductance. Therefore, in order to achieve high thermal isolation, leading to high sensitivity in photothermal detectors, it is necessary to minimize radiative heat transfer. This can be achieved by choosing a resonator material with the very low emissivity $\varepsilon$. Amorphous dielectrics, such as thin-film silicon nitride are excellent candidates with a low emissivity of the order of $\varepsilon \approx 0.05$ for typical $\sim$50 nm-thin structures.\cite{piller2020thermal,zhang2020radiative}

The second factor in equation (\ref{eq:NEP}) guiding the NEP is related to the fractional frequency ($y$) noise. Consider the common case for which electronic readout noise is negligible. Really, the NEP is the sum of uncorrelated fractional-frequency noise sources based on thermomechanical $S_{y_{thm}}$ and temperature fluctuations $S_{y_{th}}$\cite{Cleland2002} 
\begin{equation}\label{eq:S_y}
    S_y(\omega) = S_{y_{thm}}(\omega)+S_{y_{th}}(\omega)\,.
\end{equation}

$S_{y_{thm}}$ results from the thermomechanical amplitude vibration of the nanomechanical resonator driven by its own thermal energy. This amplitude noise, in turn, manifests itself in the frequency noise via amplitude-to-phase noise translation. For the optimal case that transduction noise is insignificant and the measurement filter bandwidth $\Delta\omega$ is smaller than the peak width ($\Delta\omega \ll \omega_0/Q$) of the resonator with negligible damping, it reduces to a white noise source with the power spectral density\cite{Demir2021} 
\begin{equation}\label{eq:S_ythm}
    S_{y_{thm}}(\omega) = \frac{1}{2Q^2}\frac{S_{z_{th}}}{z_r^2}\,,
\end{equation}
where $Q$ is the quality factor, $z_r$ is the vibrational amplitude of the resonator, and $S_{z_{th}}$ is the power spectral density of the thermomechanical displacement noise at the resonance frequency. Equation (\ref{eq:S_ythm}) shows that the influence of thermomechanical noise can be minimized by driving the nanomechanical resonator to the maximal vibrational amplitude limit, at the onset of nonlinearity.

$S_{y_{th}}$ comes from thermostatistical fluctuations of the resonator temperature according to the fluctuation-dissipation theorem, which directly produces frequency noise. For a lumped-element model with a concentrated mass linked to a thermal reservoir via a thermal conductance $G$, it can be simplified to \cite{vig1999noise,Cleland2002}
\begin{equation}\label{eq:S_yth}
    S_{y_{th}}(\omega) = \frac{4 k_B T^2}{G} \frac{1}{1+\omega^2\tau_{th}^2} [\mathscr{R}_T(\omega)]^2\,. 
\end{equation}

%[RW] verify the below statement and consider rewording.
In the case that $G$ is dominated by conductive heat transfer, the origin of $S_{y_{th}}$ can be explained as temperature fluctuations due to the resistance in the conductive heat transfer. In the other case that the resonator is fully in the radiative heat transfer regime, the origin of $S_{y_{th}}$ can be understood as temperature fluctuations created by the statistical nature of emitted and received thermal photons. 
Both the thermomechanical and temperature fluctuation noise can be potentially suppressed by cooling.
\cite{Zhang2023demonstration}

In the special case that heat transfer by conduction is negligible, for structures with sufficient surface area such that heat transfer due to radiation dominates ($G\approx G_{rad}$), thermal frequency fluctuations will also likely dominate ($S_{y_{thm}} \ll S_{y_{th}}$). As a result, the NEP (\ref{eq:NEP}) with (\ref{eq:R_P}) and (\ref{eq:S_yth}) reduces to
\begin{equation}
    \text{NEP} = \sqrt{4 k_B T^2 G}\,.
\end{equation}
This expression shows that NEP can be optimized by minimizing $G$ and by lowering the temperature of the resonator. However, the exact prediction of NEP remains uncertain as both the thermomechanical \cite{Sansa2016, Sadeghi2020a, Demir2021} and temperature fluctuations \cite{Zhang2023demonstration} remain an active subject of investigation.

\subsection{Sensitivity Limit of a String Resonator}
Certain nanomechanical resonators are capable of single molecule and single particle spectroscopy and microscopy. Their expected ultimate sensitivity limit is characterized by the NEP, described in (\ref{eq:NEP}). To understand its dependencies, the contributing noise mechanisms in (\ref{eq:S_y}) have to be dissected. For simplicity, we consider the sensitivity limit of a string resonator estimated by the interplay of fundamental thermophysical noise mechanisms rooted in the 
equipartition theorem. A nanostring resonator will have the parameters of length $L$, a cross-sectional area $A$, a mass density $\rho$, a pre-stress $\sigma$, and a Young's modulus $E$.

The first phenomenon contributing to the NEP, the thermomechanical fractional frequency noise (\ref{eq:S_ythm}) at resonance comes to \cite{Schmid2023}
\begin{equation}
    S_{z_{th}} = \frac{4 k_B T Q}{m_{eff}\omega_0^3}\,.
\end{equation}
Here, $m_{eff}$ is the effective mass of the resonator with the eigenfrequency $\omega_0 = (n\pi/L) \sqrt{\sigma/\rho}$ for a specific mode $n$.

As discussed above, $S_{y_{thm}}$ can be minimized by maximizing the coherent vibrational amplitude $z_r$ of the resonator. The string resonator's geometrical nonlinearity determines the maximal amplitude, given by\cite{lifshitz2008nonlinear}
\begin{equation}
    z_r \approx  1.24 \sqrt{\frac{m_{eff} \omega_0^2}{Q \alpha_{eff}}}\,,
\end{equation}
where $\alpha_{eff}$ is the effective Duffing nonlinearity parameter, which for a string is given by $\alpha_{eff} = (n \pi)^4 E A / (8 L^3)$.\cite{Schmid2023}

A second, separate, thermal phenomenon is the thermal fluctuation fractional frequency noise (\ref{eq:S_yth}), which scales inversely with the thermal conductance $G = G_{cond} + G_{rad}$, comprising conductive and radiative heat transfer. This model is for a lumped-element system, which requires the use of an effective thermal conductance $G^*$ for the string. Since thermal fluctuations happen all along the length of the string, the effective $G_{cond}^*$ can be derived by averaging the thermal resistance over the entire string length $L$
\begin{equation}
    \frac{1}{G_{cond}^*} = \frac{1}{\kappa A}\frac{1}{L}\int_0^L{\left[ \frac{1}{x}+\frac{1}{L-x}  \right]^{-1}}\text{d}x =  \frac{L}{6\kappa A}\,,
\end{equation}
where $\kappa$ is the thermal conductivity of the string's material.
The conductance due to thermal radiation ($G_{rad}^*$) is given by the Stefan-Boltzmann law. Accounting for radiation of both the top and bottom surfaces of the string and assuming a linear temperature field,\cite{Schmid2023} the total effective thermal conductance becomes
\begin{equation}
G^* = \frac{6\kappa{A}}{L} + 4 L w \sigma_B \varepsilon T^3\,,
\end{equation}
where $\sigma_B$ is the Stefan-Boltzmann constant, $\varepsilon$ is the emissivity, and $w$ the string width.

\begin{figure}
\centering
\includegraphics[width=0.4\textwidth]{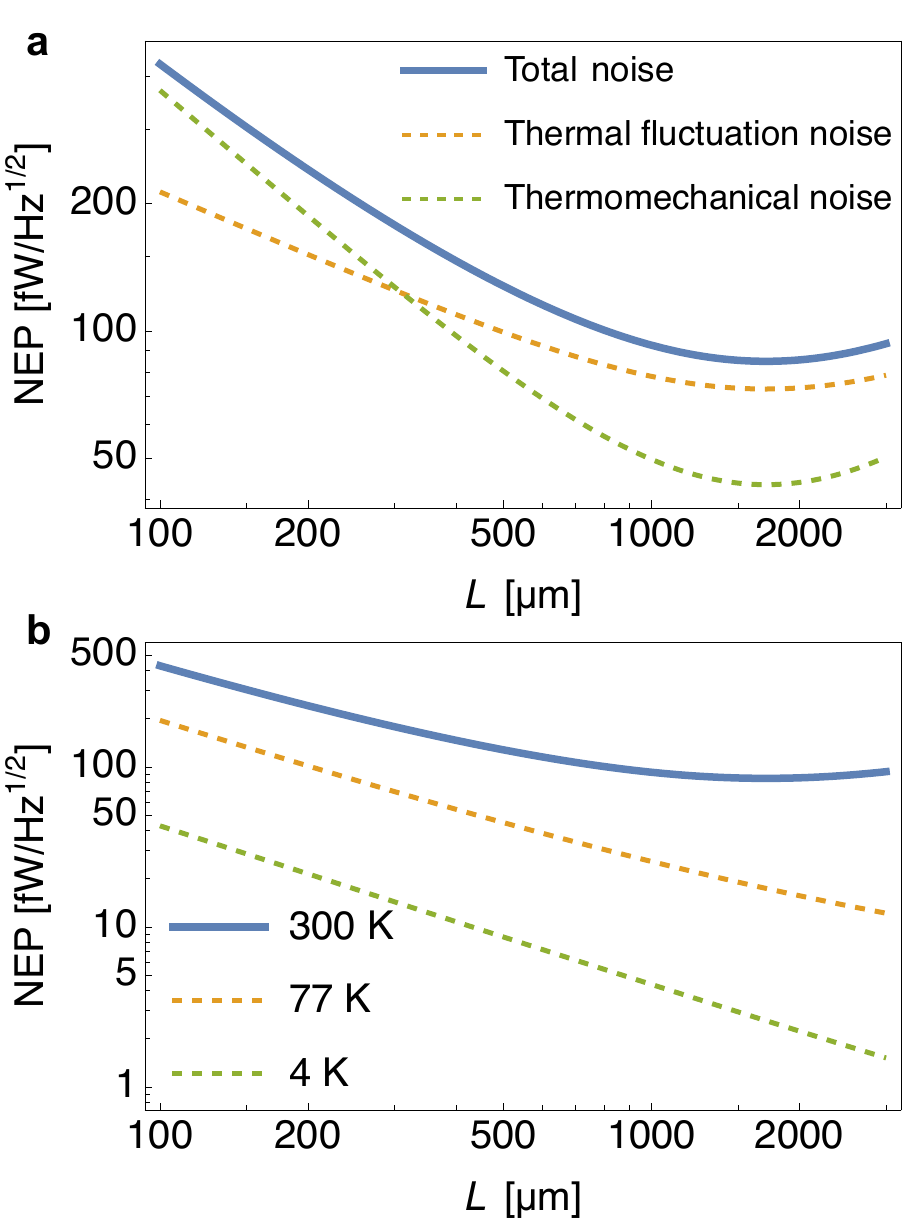}
\caption{The noise equivalent power (\ref{eq:NEP}) calculated for the fundamental mode of a 50~nm thick and 1~µm wide  SiN string resonator for various lengths, based upon the interplay of the two limiting thermal noise mechanisms (a) and the profiles for resonators cooled by liquid hydrogen and helium compared to room temperature (b). The plots are based on the following parameters: $\rho$ = \qty{2700}{\kg/\metre^3}, $E$ = \qty{250}{\GPa}, $\kappa$ = \qty{3}{\W/(\m \K)}, $\alpha$ = \qty{2.2}{ppm/\K}, 
 $\epsilon_{rad}$ = 0.05.}
\label{fig:NEP-limit}
\end{figure}

An "interplay" of these two phenomena contributing to the NEP with differing string lengths is modeled for typical SiN 50 nm-thick resonators in Figure \ref{fig:NEP-limit}a. It can be seen that for longer strings, at room temperature the thermomechanical noise is not ubiquitously the primary noise source, as often suspected\cite{Sansa2016}. For longer strings, thermal fluctuations constitute the frequency noise limit with a minimal $NEP \approx$ \qty{100}{\femto\W/\Hz^{1/2}} at room temperature. The temperature plays a major role in this sensitivity as well, yielding NEP values of just a few femtowatt at $\sim$4~K for longer strings (seen in Figure \ref{fig:NEP-limit}b). Thermal fluctuation noise has a quadratic temperature dependence and becomes negligible for temperatures below room temperature and thermomechanical noise becomes the limit. In regard to sensitivity, within these dependencies and parameters are hidden the full breadth and capability of the nanomechanical photothermal spectroscopy method.

In summary, the fundamental thermally-generated fluctuation limitation of nanoscale resonators demarcates the incomparable sensitivities of NEMS resonators to temperature-dependent changes in stress. This requires a careful implementation of the transduction of a resonator's motion to retain this fundamental limit. That is, the frequency noise in the process of transduction must be lower than fluctuation dissipation noise. Then, further improvements in sensitivity rely upon reducing the factors which govern the thermal fluctuation noise. The performance of dynamic applications will inherently require striking a balance between the speed of the resonator and its sensitivity, as these two factors are inversely proportional.

\section{Future Applications}
From thin films, to surfaces, and down to single molecule spectromicroscopy, nanomechanical resonators possess wide-ranging capability still to be explored. Immediate applications include characterization of 2D materials\cite{Kirchhof2022,Kirchhof2023} and various plasmonic structures,\cite{Chien2021,Rangacharya2020}. Considering nanomechanical resonators can operate in UHV and at cryogenic temperatures potentially places these detectors at the forefront of other fields of research. Though applications in cutting-edge heterogeneous catalysis, for example, and quantum systems are on the horizon, the study of functional nanomaterials will continue to be consequential, a springboard into these fields. Even interactions on materials in thermal contact with the resonator, such as thin layers or nanoparticles, may also be observed. Principally, this method is susceptible to all forms of spectroscopy, which can be performed in free-space. As the polarization-dependence has already been shown for single gold nanorods,\cite{Chien2020, kanellopulos2023} one immediate example is circular dichroism spectroscopy of surface-adsorbed molecules or particles, especially relevant to pharmacology.  Other applications, mentioned here are clearly within our reach, outlining a rich and exciting future for nanomechanical absorption spectroscopy.

\subsection{Analysis of functional nanomaterials} 
Functional nanomaterials, such as luminescent nanoparticles and nano-scale catalysts, are some of the most consequential advancements of nanotechnology. For example, quantum dots have emerged at the forefront of many imaging, sensing, and photonic applications. Moreover, metal nanoparticles and nanoclusters not only allow for higher catalytic efficiency but enable deeper exploration into the rich landscape of heterogeneous catlaysis.\cite{ERC_Liu_2018} On the single-particle level, particles from the same batch are most certainly heterogeneous in their size, shape, and surface characteristics. This is a crucial concern in fundamental research related to materials chemistry and surface science. In particular, for atomically precise metal nanoclusters, differences of a single atom significantly affect catalytic reaction mechanisms and optical properties.\cite{ERC_Garcia_2020} Furthermore, it is often challenging to scale up nanomaterial synthesis, and characterization of such materials is only possible on the scale of a single particle.

Single-particle microscopy, such as atomic-force microscopy or electron microscopy, has been crucial in understanding the heterogeneity of nanoparticles and has facilitated functional materials development.\cite{ERC_Zhou_2020} Likewise, specific optical properties of single nanoparticles are a vital feature in many applications. For metal species, optical properties are strongly correlated with particle size and structure. The corresponding absorption bands are located in the UV-Vis and will generally shift to a lower wavelength for decreasing particle size.\cite{ERC_V_zquez_V_zquez_2009} Therefore, UV-Vis spectroscopy is an essential characterization method for synthesized metal nanoparticles.\cite{ERC_Liu_2018,ERC_V_zquez_V_zquez_2009} But because standard UV-Vis spectroscopy can only measure the average absorption signal of the particle ensemble, the sample has to be separated and purified with, e.g., size-exclusion chromatography (SEC).\cite{ERC_Truttmann_2020,ERC_Pollitt2020} Whereas, nanomechanical photothermal spectroscopy is capable of enabling the distinction between individual particles' distinctive features and provide direct information about sample heterogeneity without separation steps. The method allows for spectroscopy across the full spectral range, where IR in combination with UV-Vis absorption can be used to gain chemical information of a nanomaterial, monitor ligand exchange reactions, if enough sample material is available \cite{ERC_Truttmann_2020}, and investigate luminescent nanoparticles as well as chiral nanomaterials.\cite{Warning2021}

Table~\ref{tab:samples} compares the previously-measured, highly fluorescent Atto 633 (with a signal-to-noise ratio of 70)\cite{Chien2018} with other single molecules and particles with their attenuation coefficients, cross-sections, and heat dissipation ratios. The single-particle sensitivity will allow the characterization of precious, scarce nanomaterials, which are difficult to fabricate, and nanoresonator spectrometers are also a robust platform for probing single thermoplasmonic and dipole-dipole interactions with their dependencies, which have a significant impact on their function in numerous applications from medicine to light harvesting.\cite{Yu2019,Mathew_2022} Metallic nanoparticles themselves have been crucial instruments in the spectroscopy of attached or neighboring analyte and have demonstrated enhancement of absorption in multiple solar cell configurations.\cite{Kundu2017} Plasmonic-enhanced spectroscopy on a thermally sensitive nanoresonator, however, will enable plasmon-enhanced absorption spectroscopy of even non-fluorescing trace substances and single molecules, improving the signal-to-noise ratio and allowing for investigations into their spectral dependencies on surface interactions. 

\subsection{Analysis of single biomolecules}
In the single-molecule sensitivity regime, a sample can be identified by way of its individual specific absorption coefficient. The amount of light a sample species attenuates at a specific wavelength is determined by its molecular composition, and can be represented by its molar attenuation coefficient $\varepsilon$. In proteins, UV light's attenuation at $\sim$280~nm can primarily be attributed to the three amino acids tryptophan (Trp), tyrosine (Tyr), and cysteine (Cys). Hence, a proteins’ molar attenuation coefficient can be estimated from the sum of the number of individual contributions of these three amino acids \cite{pace1995measure}
\begin{equation}
\begin{split}
\varepsilon_{280} [M^{-1} cm^{-1} ] &\approx (\text{\#Trp)}\times5500\\ 
\cdots&+(\text{\#Tyr})\times1490\\ 
\cdots&+(\text{\#Cys})\times125.
\end{split}
\end{equation}
The molar attenuation coefficient is highly specific to the amino acid composition of an individual protein.\cite{pace1995measure} Hence, the UV absorption from a single protein can be used as a molecule-specific fingerprint, much in the same way the mid-IR spectral fingerprint is used to identify a myriad of molecular species.

IR bands not only serve an identification function, but a quantification function for proteins. It has been shown that the attenuation signal at the amide I band (6 µm) correlates strongly with the amino acid count of a protein.\cite{strug2014development,de2021amino} The amide I signal is created by the sum of all peptide bonds that link consecutive amino acids. Therefore, from the amide I attenuation of an individual protein, it is possible to estimate the number of amino acids and identify the protein.

Concerted effort is being directed toward applications such as proteomics to find alternatives for protein identification with improved sensitivity.\cite{timp2020beyond} Acquiring chemical fingerprints, in addition to mass, allows for a multi-physical analysis, which provides an enhanced knowledge base for the identification and characterization of individual proteins and protein complexes. Here, nanomechanical single-molecule UV-Vis combines with single-molecule IR absorption spectroscopy to create a promising new method for protein identification and biomolecular analysis, in general.

\begin{table*}[]
\begin{tabular}{lccccc}

& \textbf{} 
& \textbf{Molar} 
& \textbf{Atten.} 
& \textbf{Heat} 
& \textbf{} 
\\
      
& \textbf{Spectral} 
& \textbf{attenuation} 
& \textbf{cross-} 
& \textbf{dissipation} 
& \textbf{} 
\\

& \textbf{wavelength} 
& \textbf{coefficient} 
& \textbf{section} 
& \textbf{ratio} 
& \textbf{Ref} 
\\
&                     
& $\varepsilon$ {[}M$^{-1}$ cm$^{-1}${]}                
& $A$ {[}m$^2${]}                 
& $\beta$                      
&     
\\
\hline\hline
Hepatitis B virus protein & UV (280 nm)         & 7,320,000                     & 2.8E-18                   & $\sim$0.8              & \citenum{porterfield2010simple}  \\ \hline
4 nm Au nanoparticle             & Vis (506 nm)        & 3,600,000                     & 1.4E-18                   & 1                      & \citenum{liu2007extinction}  \\ \hline
BSA protein                      & IR (6 µm)           & 190,000                           & 7.3E-20                   & 1                      & \citenum{schwaighofer2021broadband}  \\ \hline
\textbf{Atto 633}       & Vis (633 nm)        & 130,000                           & 5.0E-20                   & 0.38                   & \citenum{Chien2018}   \\ \hline
BSA protein                      & UV (280 nm)         & 44,000                            & 1.7E-20                   & $\sim$0.8              & \citenum{pace1995measure}  \\ \hline
Lysozyme protein                 & UV (280 nm)         & 38,000                            & 1.5E-20                   & $\sim$0.8              & \citenum{pace1995measure}  \\ \hline
Lysozyme protein                 & IR (6 µm)           & 36,000                            & 1.4E-20                   & 1                      & \citenum{schwaighofer2021broadband}  \\ \hline
Hepatitis B virus monomer        & UV (280 nm)         & 30,500                          & 1.2E-20                   & $\sim$0.8              & \citenum{porterfield2010simple}  \\ \hline
Virus RNA nucleotide             & UV (260 nm)         & 8,000                             & 3.0E-21                   & $\sim$1                & \citenum{porterfield2010simple}  \\ \hline
Insulin protein                  & UV (280 nm)         & 6,000                             & 2.3E-21                   & $\sim$0.8              & \citenum{pace1995measure}  \\ \hline
Tryptophan (amino acid)          & UV (280 nm)         & 5,500                           & 2.1E-21                   & 0.8                    & \citenum{pace1995measure}  \\ \hline
Virus RNA nucleotide             & UV (280 nm)         & 4,000                             & 1.5E-21                   & $\sim$1                & \citenum{porterfield2010simple}  \\ \hline
Tyrosine (amino acid)            & UV (280 nm)         & 1,490                          & 5.7E-22                   & 0.86                   & \citenum{pace1995measure}  \\ \hline
Single peptide bond              & IR (6 µm)           & 312                           & 1.2E-22                   & 1.0                    & \citenum{rahmelow1998infrared}  \\ \hline
Cysteine (amino acid)            & UV (280 nm)         & 125                           & 4.8E-23                   & N/A                    & \citenum{pace1995measure}  \\ \hline
\end{tabular}
\caption{List of molar attenuation coefficients and cross-sections of selected samples at application-relevant wavelengths in the UV-Vis and IR. The heat dissipation ratio for proteins\cite{teale1957ultraviolet,ghisaidoobe2014intrinsic} and nucleotides\cite{peon2001dna} in the UV has been estimated from their quantum yield. The molar attenuation coefficient and cross-section are directly linked by $A=3.82\times{10}^{-27}\varepsilon$.\cite{lakowicz2006principles}}
\label{tab:samples}
\end{table*}

\subsection{Fingerprinting and separation science}
 The published limits of detection for non-fluorescent analyte in the range of pg to fg of non-fluorescent analyte testify to the high sensitivity of the nanomechanical photothermal spectroscopy method. Fundamentally, measurement of such trace amounts can be accomplished with unfocused light, which minimally requires a single optical element, attesting to the simplicity of its implementation. Although, nanomechanical resonators function only in low-pressure environments, they are naturally well-suited for applications in surface science. They have already proven to be an effective way to analyze and monitor thin films and their phase transitions.\cite{Karl2020,Samaeifar2019} Furthermore, spectral "eavesdropping" can forseeably reveal information about a molecule's interactions with the surface of the resonator or vicariously through a nanoparticle on its surface. This includes monitoring of isomeric transitions and binding affinities of trace substances down to the single molecule level, adding valuable insight into surface reaction processes. This method may also avail itself to surface-assisted laser desorption of analyte,\cite{Abdelhamid_2016,Huang2022} which adsorbs directly on the mechanically resonant material on another layer placed on its surface. In this way, the influence of individual chemical groups of a chemical species can be monitored spectrally during isothermal desorption while being probed with narrow-linewidth IR light, for example. At visible wavelengths, scanning focused light would potentially enable surface sampling mass spectrometry imaging directly on the resonator.\cite{Huang2022,Wong2023} After all, the method provides time-dependent spectral data in harmony with mass exchange, which carries with it valuable information regarding trends in desorption energies, as demonstrated in Luhmann et al.\cite{Luhmann2023}. This work also demonstrates that, in mixtures, species can be differentiated and identified simply by singular value decomposition or by global spectro-temporal and, potentially, principal component analyses during isothermal desorption. This avenue of application leaves much yet to be explored but reveals a unique direction for separation science, where temperature control and chemical composition of the surface can be adapted for the separation of diverse analyte mixtures.
 
\section{Challenges}
 However, as with many technologies at the cutting edge of research, real-world implementation remains the primary challenge in regard to commercial and industrial applications for trace substance identification and quality control. The primary challenge lies in the impediments of vacuum systems; they are bulky and have some pump-down time and venting time. Nonetheless, the trade-off between processing time and sensitivity does not matter for methods such as when nanomechanical IR absorption spectroscopy of thin films was compared to ATR-FTIR measurements.\cite{Andersen2016} At other times, conventional methods introduce artefacts in the spectra which are not seen in nanomechanical photothermal spectroscopy, and the vacuum environment will be a necessary trade-off, especially at the single-molecular level.\cite{CasciCeccacci2019} Nonetheless, for trace substance analysis, with an apt system for sample exchange, the speed at which measurements are performed can be improved. Alternatively, at the expense of optimal sensitivity due to ballistic losses, many first-stage pumps can reach an ultimate partial pressure of one order of magnitude beyond $\sim10^{-3}$ mbar in air, which can be reached in a few minutes.\cite{You2021} At these pressures, however, a stabilized vacuum or a compensation mechanism for frequency drift is required.

 With the potential for sub-nanometer localization resolution at visible wavelengths, nanomechanical photothermal spectroscopy shows potential to resolve the whole range of particulate matter, which threatens global health and the ecological environment,\cite{Ault2017} and spectromectrically distinguishing between surface-adsorbed particulates.\cite{Gottschalk2023} This requires effective and accurate sampling mechanisms and sample-to-measurement tracking. Such accounting is necessary for all cases where native concentrations are desired, as analyte can be lost either in the process of sampling on resonator chips or between this and the measurement process. At low pressures, where temperature control of the chip cannot be used to mitigate loss of analyte, rates of desorption from remaining analyte after pump-down can be used to extrapolate to a near-expected initial concentration on the resonator.\cite{Shakeel2018,Luhmann2023}

\section{Conclusions}
The nanomechanical photothermal spectroscopy method provides a means to measure thermal relaxation of electronically and vibrationally excited substances down to the single molecule level. The substance being measured, whether as a thin layer, adsorbed molecule, or a suspended material, becomes part of the detection mechanism itself. Nanoresonators' versatility in fabrication, design, sampling method, and probing light characteristics has been expanding the definition of photothermal spectroscopy beyond that conventionally associated with the opto-thermal effect and merits further explorations into trace species identification, homogeneous characterization, and numerous, real-time, simultaneous or hyphenated studies. To date, studies have demonstrated that thin ceramic and 2D material resonators are capable of photothermally probing by an array of light sources and techniques: from electronic absorption of a single particle by directed, incoherent light to focused light with sub-diffraction-limited localization of nanoparticles and even FTIR of highly-dilute nebulized, aerosol-impacted solutions. Geometry and surface structure variability, along with the optical simplicity, makes this method highly adaptable. Though measurements must be performed \textit{in vacuo}, this does not necessarily mean that throughput must always be sacrificed due to pump-down and venting times. Furthermore, the thermostatistically-limited sensitivity of nanoresonators allows for accuracy beyond contemporary fingerprinting methods, and cutting edge research in surface sciences and heterogeneous catalysis requires low-pressure environments, for which spectroscopy by nanoresonators is ideal. As nanomechanical photothermal spectroscopy is inevitably leading toward single-molecule spectroscopy, there is, likewise, room for increased sensitivity where 2D materials and phononic crystals are concerned. Nonetheless, in the gap between the single molecule limit and the current state-of-the-art remains a wealth of ground-breaking applications waiting to be explored and perfected.

%\bibliography{WESTLib2,refs,ERC_CoG_2021_Citations,KKLib}
\bibliography{main.bbl}

\end{document}